\documentclass{article}

\usepackage{arxiv}

\usepackage[utf8]{inputenc} 
\usepackage[T1]{fontenc}    
\usepackage{hyperref}       
\usepackage{url}            
\usepackage{booktabs}       
\usepackage{amsfonts}       
\usepackage{nicefrac}       
\usepackage{microtype}      
\usepackage{lipsum}
\usepackage{graphicx}
\usepackage{amsmath}
\usepackage{biblatex}
\graphicspath{ {./images/} }

\title{Bifurcations in the Herd Immunity Threshold for Discrete-Time Models of Epidemic Spread}

\author{
 Sinan A. Ozbay$^1$ \\
  $^1$Bendheim Center for Finance\\
  Princeton University, USA\\
  \texttt{sozbay@princeton.edu} \\
   \And
 Bjarke F. Nielsen$^{2,3}$ \\
  $^2$PandemiX Center\\
  Roskilde University, Denmark\\
  $^3$Niels Bohr Institute \\
  University of Copenhagen, Denmark \\
  \texttt{bjarkenielsen@nbi.ku.dk} \\
  \And
 Maximilian M. Nguyen$^{4*}$ \\
  $^4$Lewis-Sigler Institute\\
  Princeton University, USA\\
  \texttt{mmnguyen@princeton.edu}\\
}

\begin{document}
\maketitle

\begin{abstract}
We performed a thorough sensitivity analysis of the herd immunity threshold for discrete-time SIR compartmental models with a static network structure. We find unexpectedly that these models violate classical intuition which holds that the herd immunity threshold should monotonically increase with the transmission parameter. We find the existence of bifurcations in the herd immunity threshold in the high transmission probability regime. The extent of these bifurcations is modulated by the graph heterogeneity, the recovery parameter, and the network size. In the limit of large, well-mixed networks, the behavior approaches that of difference equation models, suggesting this behavior is a universal feature of all discrete-time SIR models. These results suggest careful attention is needed in both selecting the assumptions on how to model time and heterogeneity in epidemiological models and the subsequent conclusions that can be drawn.
\end{abstract}

\section*{Introduction}
The herd immunity threshold (henceforth, HIT) is a key epidemiological quantity that signifies the end of the growth phase of an epidemic. During the COVID-19 pandemic, the HIT has both received a great deal of attention from scientists, media, and the public, as well has been a significant source of contention amongst policymakers \cite{fontanet_covid-19_2020, randolph_herd_2020, morens_concept_2022}. Despite its importance, the behavior of this quantity beyond classical epidemiological models is still not well understood. How the HIT is affected by heterogeneity in the spacing of infections in time and heterogeneity in contact structure remain open areas of research.

In constructing epidemic models, a key conceptual decision is deciding on how to represent the flow of time, which can be done in either a continuous or discrete fashion. For continuous-time models of epidemic spread, the underlying premise is that contact events or potential infection events are well-mixed and evenly distributed through time. Typically, this assumption is formalized as contact events being a Poisson-distributed transmission process. This is a common and convenient assumption which is used both in classical differential equation-based models, as well as in graph network models \cite{van_mieghem_virus_2009, valdano_epidemic_2018, chen_discrete-time_2019}. The assumption is also more amenable to mathematical analysis techniques.

In contrast, there are arguments that have been posed in the literature on the utility of a model of infections that is fundamentally discrete-time \cite{diekmann_discrete-time_2021, brauer_discrete_2010, hernandez-ceron_discrete_2013, block_social_2020}. Firstly, since case data (both prevalence and incidence) are typically reported discretely on a daily or weekly basis \cite{cdc_covid_2020}, it is worth investigating the consequences of models that update time on the same time-scale as the real-world data. Secondly, human mobility patterns have been empirically shown to be quite regular and concentrated in time \cite{cho_friendship_2011, song_limits_2010}, suggesting that social contact peaks at certain times during the day. These bursts of social activity are opportunities for infection that are more concentrated than a Poisson process would suggest. Finally, from a computational standpoint, it is sometimes necessary and easier to deal with discrete-time models.

It is an open question whether these two representations of time significantly differ in their behavior and predictions for key epidemic quantities, such as the HIT. 
While we do not believe time to actually be discrete in reality, there are valid reasons a modeller may choose either type of time dynamic. Thus we explore the understudied assumption of modeling time as discrete and the resulting implications of that assumption as it relates to the HIT. Any resulting differences in the behavior of the HIT must be taken as a consequence of the modeling of time, rather than necessarily a reflection of physical reality. As will be shown, a seemingly innocuous choice on how to represent time can have rather dramatic downstream consequences, serving as an important reminder to all modelers to consider their basis of assumptions meticulously. 

Another key choice in modeling epidemics is establishing the structure for how infections will spread. Realistic models of disease spread are critical for accurate forecasting and management of epidemics, as the COVID-19 pandemic has made clear. Classical models of disease spread, which have been very useful for general understanding of epidemic phenomena and for their analytical tractability in certain cases, can make unrealistic modelling assumptions about the structure of the underlying contact network. This simplification can render these models unsatisfactory as tools for predicting the extent and severity of disease spread. For instance, the classical SIR compartmental model assumes a contact network structure wherein each individual is (1) connected to all other individuals in the network and (2) the size of the network approaches a magnitude that allows for stochastic effects of spread to be ignored. Often, neither of these assumptions is realistic. One consequence is that these assumptions do not mathematically allow for known epidemic phenomena, such as superspreaders \cite{lloyd-smith_superspreading_2005, chang_mobility_2021, nielsen_covid-19_2021}, to occur. Recently, the explosion of computational resources for simulation and availability of granular data for contact network estimation have made agent-based models that simulate spread of epidemics on realistic networks a viable and practical tool \cite{eubank_modelling_2004, pastor-satorras_epidemic_2015}. Recent work has also begun to explore the impact of heterogeneity on HIT \cite{fine_herd_2011, britton_mathematical_2020, keeling_networks_2005, ferrari_network_2006, metcalf_understanding_2015}.

To investigate how these modeling choices influence the HIT, we considered simulations of agent-based models that use graphs of varying degree heterogeneity with infection events occurring in discrete time. As a result, the states of all individuals in the network are updated simultaneously at each time-step. This relaxes the above assumption that infections occur uniformly through continuous time. It is the purpose of this paper to shed light on the highly non-classical behaviors these models can generate and subsequently discuss the implications for epidemic forecasting.

\section*{Results}
Results are reported for the following numerical experiments. We conducted simulations of SIR epidemic spread on an agent-based model, where individuals infect contacts on a network structure represented by a graph. We performed a thorough sensitivity analysis on four critical model parameters: the network heterogeneity parameter $\sigma \in [0,1]$, which controls the heterogeneity in the degree-distribution of the network generated \cite{ozbay_parameterizing_2022}; the transmission probability $\tau$, which determines how likely an infected individual is to infect a susceptible neighbor in a given time-step; the spontaneous recovery parameter $\gamma$, which controls how probable it is that an infected individual recovers in a given time-step, and the size of the network $n$. 

Our simulations generate time-series of the number of individuals in a given compartment: Susceptible, Infected, or Recovered, from which we compute key epidemic statistics: the HIT and the time at the peak of infection prevalence. For ODE-based SIR models, the HIT can be derived analytically in terms of the basic reproduction number $R_0$ \cite{anderson_vaccination_1985}. Since $R_0$ does not have a standard analogue in the context of epidemic spread on a network, we have defined the HIT based on a key feature on when it would classically occur, namely that the HIT occurs at the peak of infection prevalence. Here we defined HIT as the cumulative proportion of the population that has been reached by the disease by the time the infection prevalence peaks. This definition is meant to intuitively align with classical expectations of a herd immunity threshold, namely that it is the point at which the effective reproductive number equals 1 and where the epidemic is no longer growing in size.

As we scan through parameter space, we study the distribution and behavior of the HIT and time to infection prevalence to understand the behavior of epidemic peaks under iteration. Although we allow the topology of the contact networks to vary, all edges carry equal weight $(\tau)$. For each unique quadruplet of parameters: transmission probability, recovery probability, graph heterogeneity, and size given as $(\tau, \gamma, \sigma, n)$, we run 150 simulations, randomly infecting a small initial fraction of nodes on the graph. Further details of the construction of the contact network and the model for simulations can be found in the Methods section.

\begin{figure}
    \centering
    \centerline{\includegraphics[scale=0.6]{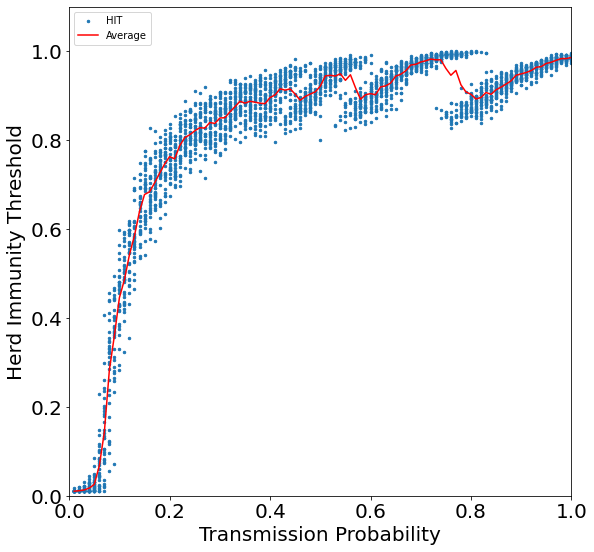}}
    \caption{The herd immunity threshold as a function of the transmission probability, $\tau$, for the classical SIR model. The remaining SIR model parameters are fixed at $\sigma \approx 0,\gamma = 0.2, n = 1000$. The epidemic is reinitialized for 150 iterations at each value of $\tau$.  Blue dots denote individual simulation runs. The red line denotes an average over simulation runs at each value of $\tau$.} \label{fig:1}
\end{figure}

In Figure \ref{fig:1}, we display the HIT as a function of transmission probability for a single $(\gamma, \sigma, n)$ triple and the full range of transmission probabilities ($\tau$). Each point represents the results of a single simulated epidemic, and the red line represents the empirical average across iterations for a single value of $\tau$. 

Starting from the lowest transmission probability and moving right, we first encounter a subcritical phase gated by a critical threshold value of $\tau$. Upon crossing the threshold, the average HIT rapidly grows before saturating at an asymptotic phase which is characterized by oscillations in the average HIT and bifurcations in the individual point clouds. In what follows, we present further results that test the sensitivity of this bifurcation behavior to various parameters of the SIR model.

\subsection*{Increased Contact Heterogeneity Suppresses Bifurcations}

We investigated the sensitivity of the HIT to the heterogeneity of the underlying contact network. The heterogeneity of the network is given by a single parameter $\sigma$ which, at $\sigma = 0$, specifies a set of graphs with completely homogeneous degree distributions (all nodes in the graph have the same degree), while $\sigma = 1$ specifies a graph whose degree distribution is heavy-tailed. More details on this parameter and the method used to generate the degree distributions is included in the Methods section. 

\begin{figure}
    \centering
    \includegraphics[width=0.32\textwidth]{Sigma1Final.png}
    \includegraphics[width=0.32\textwidth]{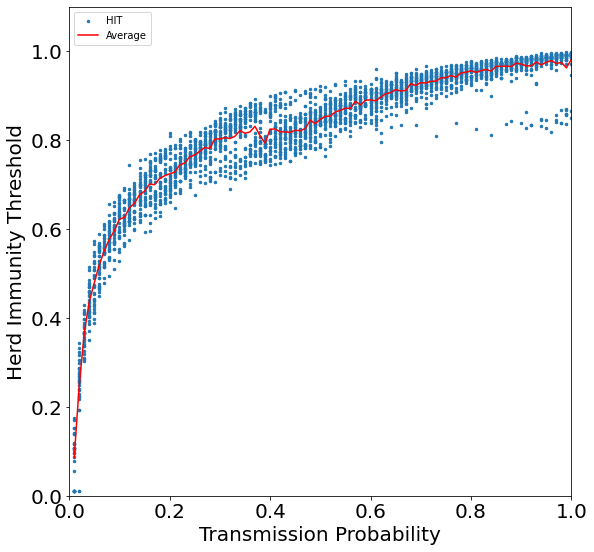}
    \includegraphics[width=0.32\textwidth]{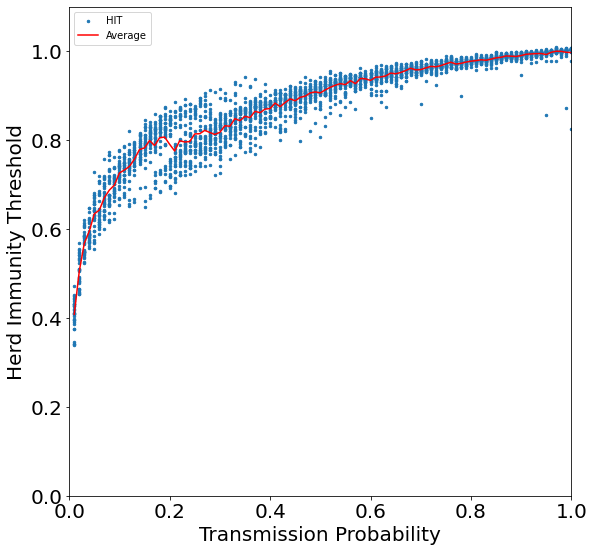}
    \caption{Herd immunity threshold as a function of transmission probability for different levels of graph heterogeneity. Left to right, the value of the heterogeneity parameter $\sigma$ is increasing: $[0, 0.5, 0.82]$. The remaining SIR model parameters are fixed at $\gamma = 0.2, n = 1000$. The epidemic is reinitialized for 150 iterations at each value of $\tau$.  Blue dots denote individual simulation runs. The red line denotes an overage over the simulation runs at each value of $\tau$.} \label{fig:hetero}
\end{figure}

In Figure \ref{fig:hetero}, it can be seen that the HIT is generally increasing as a function of the transmission probability. More homogeneous graphs have a subcritical regime where the HIT and epidemic does not grow to an appreciable size until a critical transmission probability threshold is reached. The disappearance of this threshold in more heterogeneous networks, particularly scale-free networks, is well documented \cite{pastor-satorras_epidemic_2001}. Increasing graph heterogeneity appears to decrease the number and the gaps in the bifurcations in the distribution of the HIT.

\subsection*{Increased Recovery Probability Drives Variability in Epidemic Peak}

\begin{figure}
    \centering
    \includegraphics[width=0.32\textwidth]{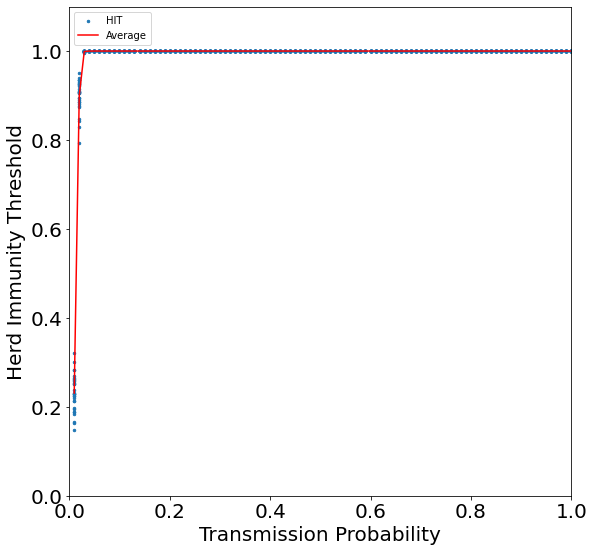}
    \includegraphics[width=0.32\textwidth]{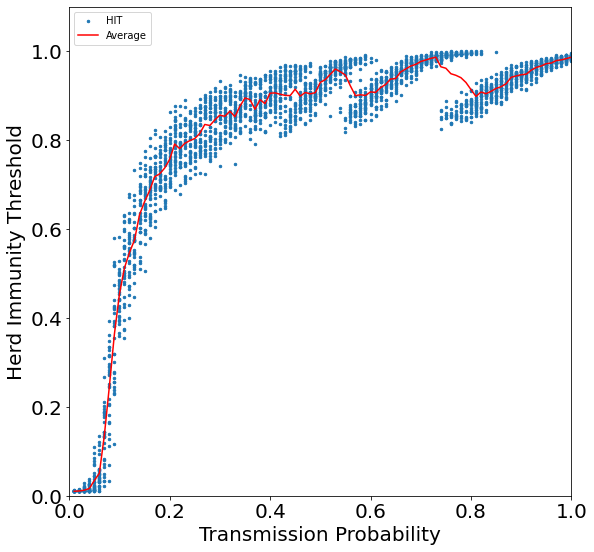}
    \includegraphics[width=0.32\textwidth]{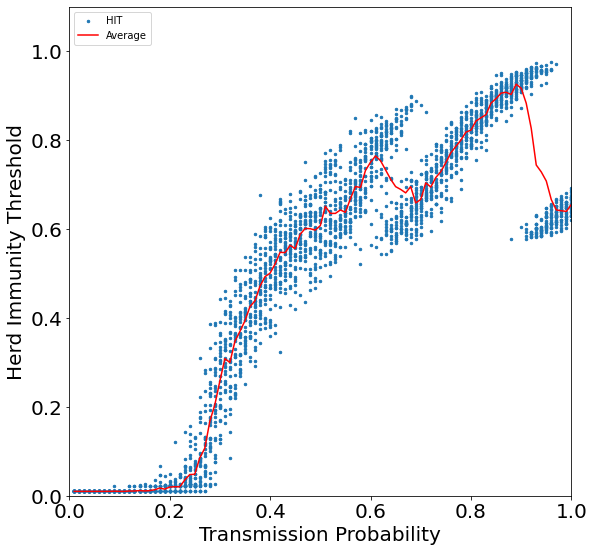}
    \caption{Herd immunity threshold as a function of transmission probability for different recovery rates. Left to right, the value of the spontaneous recovery parameter $\gamma$ is increasing: $[0, 0.2, 1]$. The remaining SIR model parameters are fixed at $\sigma \approx 0, n = 1000$. The epidemic is reinitialized for 150 iterations at each value of $\tau$.  Blue dots denote individual simulation runs. The red line denotes an overage over the simulation runs at each value of $\tau$.} \label{fig:recovery}
\end{figure}

We investigated the sensitivity of the HIT to the spontaneous recovery probability $\gamma$. In Figure \ref{fig:recovery}, it can be seen that as $\gamma \to 0$ (which is the SI limit of the SIR model) there is no bifurcation in the HIT observed and convergence to a completely infected network ($HIT = 1$) is very rapid, as there is no recovery. As we increase the recovery probability, the convergence to the asymptotic phase is slower as a function of transmission probability. In addition, the extent of the subcritical phase becomes elongated for higher recovery probabilities. The range of potential HIT values also increases with increasing recovery probability, allowing for significantly lower HIT values to be possible. Increasing the recovery probability also increases the magnitude of the gaps in HIT bifurcations in the asymptotic phase. 

\subsection*{Bifurcations Sharpen and Increase in Number with System Size}

\begin{figure}
    \centering
    \includegraphics[width=0.32\textwidth]{Sigma1Final.png}
    \includegraphics[width=0.32\textwidth]{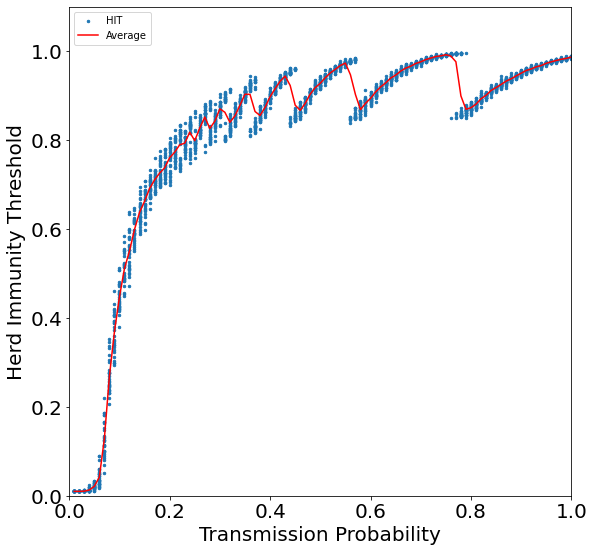}
    \includegraphics[width=0.32\textwidth]{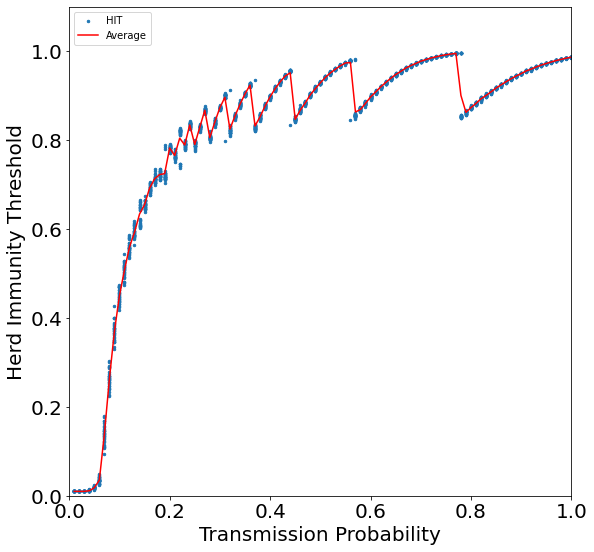}
    \caption{Herd immunity threshold as a function of transmission probability for different network sizes. Left to right, the size of the graph on which the epidemic is simulated, $n$, is increasing: $[1000, 10000, 100000]$. The remaining SIR model parameters are fixed at $\sigma \approx 0, \gamma = 0.2$. The epidemic is reinitialized for 150 iterations at each value of $\tau$.  Blue dots denote individual simulation runs. The red line denotes an overage over the simulation runs at each value of $\tau$.} \label{fig:size}
\end{figure}

We considered the effect of changing the size of the system on the HIT by changing the number of nodes on which we simulate the epidemic. In Figure \ref{fig:size}, increasing the order of magnitude of the nodes leads to a greater number and frequency of bifurcations in the HIT. Additionally, the variance around the average HIT value for a given transmission probability decreases as the size of the graph increases. 

\begin{figure}
    \centering
    \includegraphics[width=0.31\textwidth]{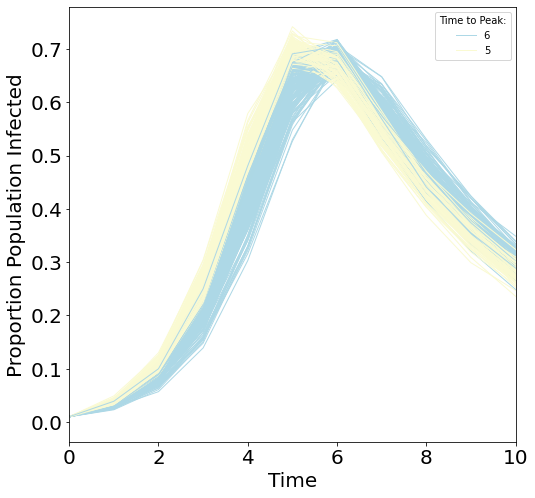}
    \includegraphics[width=0.31\textwidth]{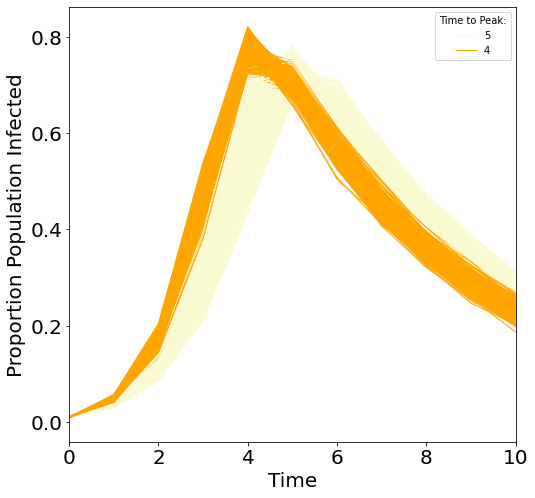}
    \includegraphics[width=0.31\textwidth]{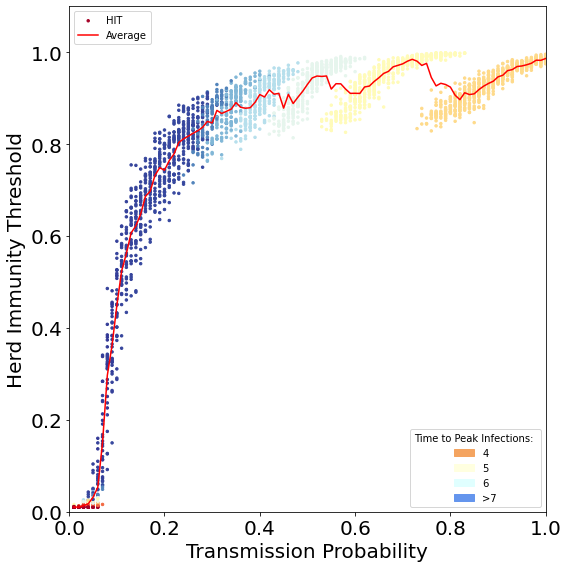}
    \caption{a) Infection curves represented as the proportion of the population infected as a function of time. SIR model parameters of $\sigma \approx 0, \gamma = 0.2, \tau = [0.5,0.6],n=1000$. Individual simulation runs are colored based on what time the peak of infections occurs. Total number of curves (individual epidemic runs) is 300. b) Infection curves represented as the proportion of the population infected as a function of time. SIR model parameters of $\sigma \approx 0, \gamma = 0.2, \tau = [0.75,0.85],n=1000$. Individual simulation runs are colored based on what time the peak of infections occurs. Total number of curves (individual epidemic runs) is 300. c) Herd immunity threshold as a function of $\tau$ with different colors indicating a different time at which the peak of infections is reached. The epidemic is repeated for 150 iterations at each value of $\tau$.  Dots denote individual simulation runs. The red line denotes an overage over the simulation runs at each value of $\tau$.} \label{fig:time}
\end{figure}

\section*{Discussion}
 The most notable of the non-classical behaviors presented is the bifurcation gaps of the HIT, as seen in Figures \ref{fig:1}-\ref{fig:size}. The classical intuition that as the transmission parameter increases, the HIT will also monotonically increase is clearly violated both at the level of individual simulations (blue dots), as well as for averages over simulations runs (red line). In certain parameter regimes, for instance Figure \ref{fig:recovery}c, the difference in HIT values across the bifurcation gap can be substantial: for a $\tau = 0.9$, the epidemic may realize an HIT of 0.6 or 1 depending on how the stochastics of the epidemic process play out. We thus observed that even the simplest models of discrete-time disease spread on networks produces behavior that is highly non-classical and even counter-intuitive by the standard results of ODE-based SIR models in the literature. 

The origin of this bifurcation behavior can be unraveled by inspecting the infection curves of the individual simulation runs as a function of time. To demonstrate, let us look at the distribution of infection curves for the interval of $\tau = [0.5,0.6]$ in Figure \ref{fig:1}, which includes a single bifurcation. We see that when we plot the infection curves for simulations with parameter values of $\tau$ in this interval as a function of time (Figure \ref{fig:time}a), all of the simulation runs segregate into exactly two groups: runs that peak at time 5 (yellow curves) and time 6 (blue curves) respectively. We note that both groups of infection curves are similar in shape, which can be verified by calculating a convolution of curves representative of the two groups, reflecting that both groups are drawn under the same parameter values. Importantly however, we note that all of the curves in yellow grow at a rate faster than the blue curves during the growth phase (times 1-4). In addition, the individual runs in the yellow group recover at a slightly faster rates, as seen by an earlier decline in infections compared to blue curves. Thus, small absolute differences in the initial spreading dynamics due to stochasticity differentiates runs into two groups, allowing some runs to peak one time-step earlier without shifting the bulk of the infection curve. While the infection curves between the two groups overlap heavily, importantly, the position of the peak shifts from one shoulder to another. Since the HIT is a cumulative summation that is sensitive to where the peak occurs, even a difference of a single time step can cause a large difference in the HIT attained between runs. Increasing the transmission probability amounts to shifting the infection curves to larger values of time, while still maintaining this dimorphism between faster-peaking runs and slower-peaking runs, as seen in Figures \ref{fig:time}b,c.

The explanation above is also consistent with the trends observed in how these bifurcations are dependent on the other parameters of the model. For instance, as the graph becomes more heterogeneous ($\sigma \to 1$) the bifurcations decrease in frequency. As heterogeneity increases, the networks become more heavy-tailed in the degree distribution. This causes the average path length between pairs of nodes to decrease as shortcuts can be taken through hubs. This is in contrast with homogeneous networks, which are characterized by a lattice-like structure and feature much longer minimum path lengths between any two nodes in the network. In effect, heterogeneity in the network allows a larger set of initial conditions and transmission probabilities to lead to a full blown epidemic. For example, in Figure \ref{fig:hetero}c the time to the peak of infection prevalence peaks quickly. This fast time to peak is the same for all runs in the entire range of $\tau = [0.15, 1]$ after the first bifurcation. There are no further bifurcations because the network can not peak any faster since the average shortest path length sets a topological lower bound on the time to peak of infections. 

The impact of the recovery rate on the phenomena can also be disentangled. In the SI limit ($\gamma \to 0$) the bifurcation behavior does not manifest, whereas in the limit of immediate recovery in the time-step following infection ($\gamma \to 1$) the bifurcation jumps are the most dramatic. Intuitively, the increased recovery parameter makes it more difficult for the epidemic to continue spreading, which allows for both lower HIT values and for critical transmission probability threshold needed for a non-trivial epidemic. Increased recovery also makes the epidemic trajectory more dependent on the initial conditions, which is reflected on the wider variation in HIT values for a given $\tau$. The effect of system size agrees with intuition that as the system size increases one should expect stochastic effects to decrease in relative magnitude. This is reflected by the decrease in smearing of data points around the average HIT with increasing system size. This is sensible, as deterministic systems should only attain a single HIT value for a given set of parameter values, given they always generate a single epidemic path.

\begin{figure}
    \centering
    \includegraphics[width=0.45\textwidth]{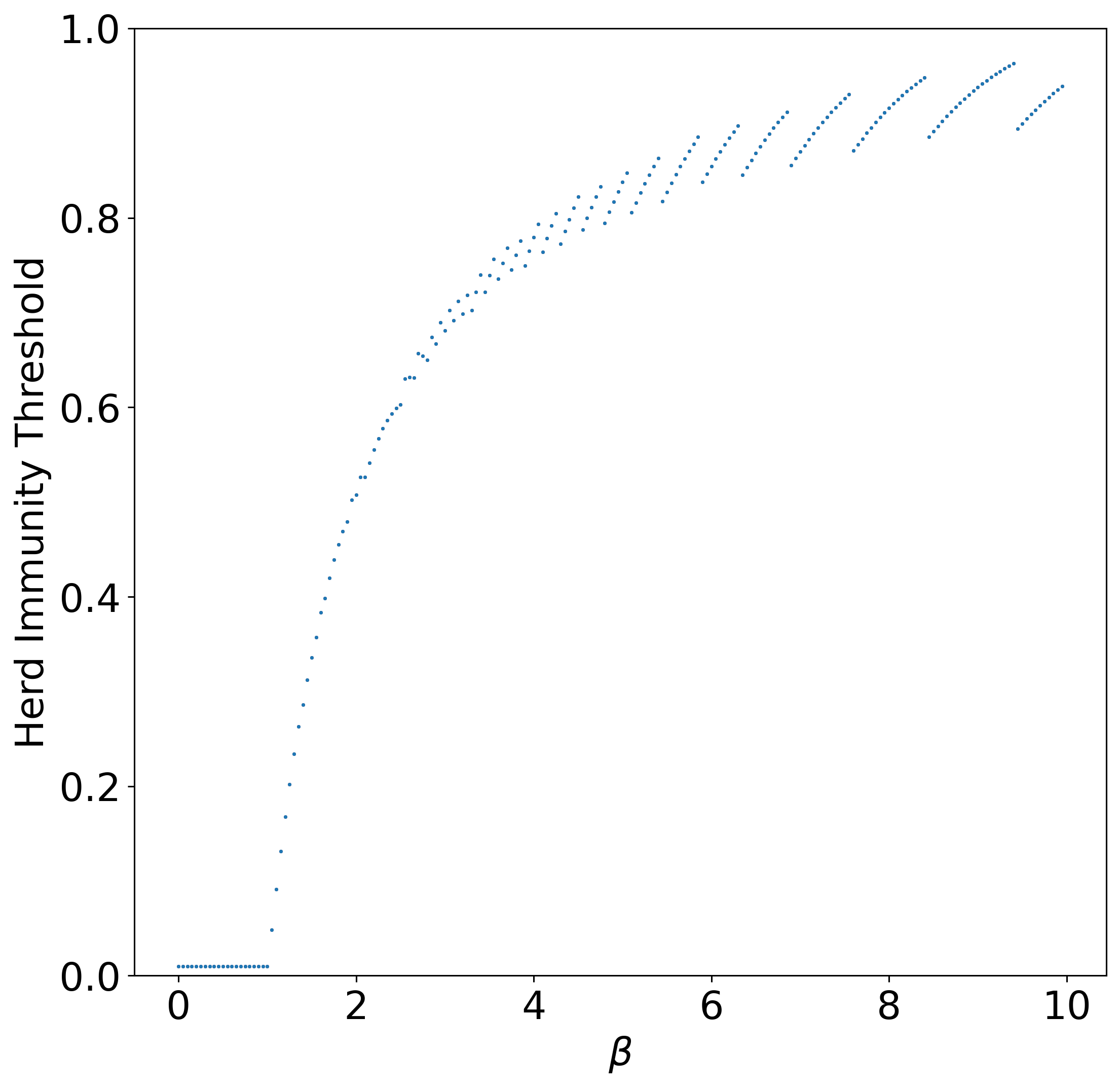}
    \includegraphics[width=0.45\textwidth]{Size3Final.png}

    \caption{a) Herd immunity threshold as a function of transmission parameter $\beta$ for the SIR model in the form of difference equations. $\beta$ represents the mean number of new infections an infected individual generates. The remaining SIR model parameter is fixed at $\gamma = 1$. Each point is an individual simulation that is run for a duration of 20 time units with a time-step size of $dt = 0.1$. b) Herd immunity threshold as a function of transmission probability. The other SIR model parameters are fixed at $\sigma \approx 0, \gamma = 0.2, n=10^5$. The epidemic is repeated for 150 iterations at each value of $\tau$.  Blue dots denote individual simulation runs. The red line denotes an overage over the simulation runs at each value of $\tau$.} \label{fig:diffeq}
\end{figure}

With the intuition given by the sensitivity analysis, we might expect then that in the large, well-mixed limit, the behavior of the HIT for network models should approach that of the Kermack-McKendrick SIR model \cite{kermack_contribution_1927}. Indeed, we found that the network behavior recapitulates the same qualitative behavior as the deterministic SIR equations under time discretization (Figure \ref{fig:diffeq}). Equations for the SIR model in the form of difference equations can be found in the Methods. This implies the bifurcation behavior of the HIT at high transmission probabilities is not restricted to SIR models that contain either network structure or stochasticity. This suggests that these bifurcations and oscillations in HIT may exist in certain regions of parameter space for all discrete-time SIR models. Other discrete-time SIR models are reported in the literature \cite{diekmann_discrete-time_2021, kreck_back_2022}. 

\section*{Conclusion}
In general, increasing the transmission probability causes the epidemic to spread more rapidly in the population, which decreases the time to peak of infections. However, for discrete-time models this is necessarily a transition from one discrete time-step to another. The incremental nature of discrete time directly yields the observed bifurcation behavior in the HIT. That the observed bifurcation behavior occurs rather generically in discrete-time SIR models suggests that a modeler is faced with a key decision when deciding on how to represent time conceptually in their models. For the reasons mentioned previously, both continuous- and discrete-time models present distinct advantages that merit their use. Here we have illustrated some key behaviors and even unintended impacts that may arise when making the choice to model time discretely and considering the role of heterogeneity. This may prove insightful not only to epidemic modelers, but also those in position to make public policy who need to understand the limits of the conclusions they can draw from the models being employed \cite{james_use_2021}. That such a simple alteration to the SIR model can lead to highly non-classical behavior suggests it would be prudent for the epidemiological community to carefully revisit and to inspect even the most basic assumptions being used in models.

\printbibliography

@article{block_social_2020,
	title = {Social network-based distancing strategies to flatten the {COVID}-19 curve in a post-lockdown world},
	volume = {4},
	rights = {2020 The Author(s), under exclusive licence to Springer Nature Limited},
	issn = {2397-3374},
	url = {https://www.nature.com/articles/s41562-020-0898-6},
	doi = {10.1038/s41562-020-0898-6},
	pages = {588--596},
	number = {6},
	journaltitle = {Nature Human Behaviour},
	shortjournal = {Nat Hum Behav},
	author = {Block, Per and Hoffman, Marion and Raabe, Isabel J. and Dowd, Jennifer Beam and Rahal, Charles and Kashyap, Ridhi and Mills, Melinda C.},
	urldate = {2022-09-22},
	date = {2020-06},
	langid = {english},
	note = {Number: 6
Publisher: Nature Publishing Group},
}

@article{chen_discrete-time_2019,
	title = {Discrete-Time vs. Continuous-Time Epidemic Models in Networks},
	volume = {7},
	issn = {2169-3536},
	doi = {10.1109/ACCESS.2019.2940132},
	pages = {127669--127677},
	journaltitle = {{IEEE} Access},
	author = {Chen, Zesheng},
	date = {2019},
	note = {Conference Name: {IEEE} Access},
}

@article{valdano_epidemic_2018,
	title = {Epidemic Threshold in Continuous-Time Evolving Networks},
	volume = {120},
	url = {https://link.aps.org/doi/10.1103/PhysRevLett.120.068302},
	doi = {10.1103/PhysRevLett.120.068302},
	pages = {068302},
	number = {6},
	journaltitle = {Physical Review Letters},
	shortjournal = {Phys. Rev. Lett.},
	author = {Valdano, Eugenio and Fiorentin, Michele Re and Poletto, Chiara and Colizza, Vittoria},
	urldate = {2022-09-22},
	date = {2018-02-06},
	note = {Publisher: American Physical Society},
}

@article{van_mieghem_virus_2009,
	title = {Virus Spread in Networks},
	volume = {17},
	issn = {1558-2566},
	doi = {10.1109/TNET.2008.925623},
	pages = {1--14},
	number = {1},
	journaltitle = {{IEEE}/{ACM} Transactions on Networking},
	author = {Van Mieghem, Piet and Omic, Jasmina and Kooij, Robert},
	date = {2009-02},
	note = {Conference Name: {IEEE}/{ACM} Transactions on Networking},
}

@article{anderson_vaccination_1985,
	title = {Vaccination and herd immunity to infectious diseases},
	volume = {318},
	rights = {1985 Nature Publishing Group},
	issn = {1476-4687},
	url = {https://www.nature.com/articles/318323a0},
	doi = {10.1038/318323a0},
	pages = {323--329},
	number = {6044},
	journaltitle = {Nature},
	author = {Anderson, Roy M. and May, Robert M.},
	urldate = {2022-09-21},
	date = {1985-11},
	langid = {english},
	note = {Number: 6044
Publisher: Nature Publishing Group},
}

@article{nielsen_covid-19_2021,
	title = {{COVID}-19 Superspreading Suggests Mitigation by Social Network Modulation},
	volume = {126},
	url = {https://link.aps.org/doi/10.1103/PhysRevLett.126.118301},
	doi = {10.1103/PhysRevLett.126.118301},
	pages = {118301},
	number = {11},
	journaltitle = {Physical Review Letters},
	shortjournal = {Phys. Rev. Lett.},
	author = {Nielsen, Bjarke Frost and Simonsen, Lone and Sneppen, Kim},
	urldate = {2022-09-21},
	date = {2021-03-15},
	note = {Publisher: American Physical Society},
}

@article{chang_mobility_2021,
	title = {Mobility network models of {COVID}-19 explain inequities and inform reopening},
	volume = {589},
	rights = {2020 The Author(s), under exclusive licence to Springer Nature Limited},
	issn = {1476-4687},
	url = {https://www.nature.com/articles/s41586-020-2923-3},
	doi = {10.1038/s41586-020-2923-3},
	pages = {82--87},
	number = {7840},
	journaltitle = {Nature},
	author = {Chang, Serina and Pierson, Emma and Koh, Pang Wei and Gerardin, Jaline and Redbird, Beth and Grusky, David and Leskovec, Jure},
	urldate = {2022-09-21},
	date = {2021-01},
	langid = {english},
	note = {Number: 7840
Publisher: Nature Publishing Group},
}

@online{cdc_covid_2020,
	title = {{COVID} Data Tracker},
	url = {https://covid.cdc.gov/covid-data-tracker},
	titleaddon = {Centers for Disease Control and Prevention},
	author = {{CDC}},
	urldate = {2022-09-16},
	date = {2020-03-28},
	langid = {english},
}

@article{pastor-satorras_epidemic_2015,
	title = {Epidemic processes in complex networks},
	volume = {87},
	url = {https://link.aps.org/doi/10.1103/RevModPhys.87.925},
	doi = {10.1103/RevModPhys.87.925},
	pages = {925--979},
	number = {3},
	journaltitle = {Reviews of Modern Physics},
	shortjournal = {Rev. Mod. Phys.},
	author = {Pastor-Satorras, Romualdo and Castellano, Claudio and Van Mieghem, Piet and Vespignani, Alessandro},
	urldate = {2022-09-16},
	date = {2015-08-31},
	note = {Publisher: American Physical Society},
}

@inproceedings{cho_friendship_2011,
	location = {New York, {NY}, {USA}},
	title = {Friendship and mobility: user movement in location-based social networks},
	isbn = {978-1-4503-0813-7},
	url = {https://doi.org/10.1145/2020408.2020579},
	doi = {10.1145/2020408.2020579},
	series = {{KDD} '11},
	shorttitle = {Friendship and mobility},
	pages = {1082--1090},
	booktitle = {Proceedings of the 17th {ACM} {SIGKDD} international conference on Knowledge discovery and data mining},
	publisher = {Association for Computing Machinery},
	author = {Cho, Eunjoon and Myers, Seth A. and Leskovec, Jure},
	urldate = {2022-09-16},
	date = {2011-08-21},
}

@article{eubank_modelling_2004,
	title = {Modelling disease outbreaks in realistic urban social networks},
	volume = {429},
	rights = {2004 Macmillan Magazines Ltd.},
	issn = {1476-4687},
	url = {https://www.nature.com/articles/nature02541},
	doi = {10.1038/nature02541},
	pages = {180--184},
	number = {6988},
	journaltitle = {Nature},
	author = {Eubank, Stephen and Guclu, Hasan and Anil Kumar, V. S. and Marathe, Madhav V. and Srinivasan, Aravind and Toroczkai, Zoltán and Wang, Nan},
	urldate = {2022-09-16},
	date = {2004-05},
	langid = {english},
	note = {Number: 6988
Publisher: Nature Publishing Group},
}

@article{metcalf_understanding_2015,
	title = {Understanding Herd Immunity},
	volume = {36},
	issn = {1471-4906},
	url = {https://www.sciencedirect.com/science/article/pii/S1471490615002495},
	doi = {10.1016/j.it.2015.10.004},
	pages = {753--755},
	number = {12},
	journaltitle = {Trends in Immunology},
	shortjournal = {Trends in Immunology},
	author = {Metcalf, C. J. E. and Ferrari, M. and Graham, A. L. and Grenfell, B. T.},
	urldate = {2022-09-16},
	date = {2015-12-01},
	langid = {english},
}

@article{keeling_networks_2005,
	title = {Networks and epidemic models},
	volume = {2},
	url = {https://royalsocietypublishing.org/doi/full/10.1098/rsif.2005.0051},
	doi = {10.1098/rsif.2005.0051},
	pages = {295--307},
	number = {4},
	journaltitle = {Journal of The Royal Society Interface},
	author = {Keeling, Matt J and Eames, Ken T.D},
	urldate = {2022-09-16},
	date = {2005-09-22},
	note = {Publisher: Royal Society},
}

@article{hernandez-ceron_discrete_2013,
	title = {Discrete epidemic models with arbitrary stage distributions and applications to disease control},
	volume = {75},
	issn = {0092-8240},
	url = {https://www.ncbi.nlm.nih.gov/pmc/articles/PMC4002294/},
	doi = {10.1007/s11538-013-9866-x},
	pages = {1716--1746},
	number = {10},
	journaltitle = {Bulletin of mathematical biology},
	shortjournal = {Bull Math Biol},
	author = {Hernandez-Ceron, Nancy and Feng, Zhilan and Castillo-Chavez, Carlos},
	urldate = {2022-09-16},
	date = {2013-10},
	pmid = {23797790},
	pmcid = {PMC4002294},
}

@article{kermack_contribution_1927,
	title = {A contribution to the mathematical theory of epidemics},
	volume = {115},
	url = {https://royalsocietypublishing.org/doi/abs/10.1098/rspa.1927.0118},
	doi = {10.1098/rspa.1927.0118},
	pages = {700--721},
	number = {772},
	journaltitle = {Proceedings of the Royal Society of London. Series A, Containing Papers of a Mathematical and Physical Character},
	author = {Kermack, William Ogilvy and {McKendrick}, A. G. and Walker, Gilbert Thomas},
	urldate = {2022-09-16},
	date = {1927-08},
	note = {Publisher: Royal Society},
}

@article{britton_mathematical_2020,
	title = {A mathematical model reveals the influence of population heterogeneity on herd immunity to {SARS}-{CoV}-2},
	volume = {369},
	url = {https://www.science.org/doi/10.1126/science.abc6810},
	doi = {10.1126/science.abc6810},
	pages = {846--849},
	number = {6505},
	journaltitle = {Science},
	author = {Britton, Tom and Ball, Frank and Trapman, Pieter},
	urldate = {2022-09-16},
	date = {2020-08-14},
	note = {Publisher: American Association for the Advancement of Science},
}

@article{diekmann_discrete-time_2021,
	title = {The discrete-time Kermack–{McKendrick} model: A versatile and computationally attractive framework for modeling epidemics},
	volume = {118},
	url = {https://www.pnas.org/doi/full/10.1073/pnas.2106332118},
	doi = {10.1073/pnas.2106332118},
	shorttitle = {The discrete-time Kermack–{McKendrick} model},
	pages = {e2106332118},
	number = {39},
	journaltitle = {Proceedings of the National Academy of Sciences},
	author = {Diekmann, Odo and Othmer, Hans G. and Planqué, Robert and Bootsma, Martin C. J.},
	urldate = {2022-09-16},
	date = {2021-09-28},
	note = {Publisher: Proceedings of the National Academy of Sciences},
}

@article{kreck_back_2022,
	title = {Back to the Roots: A Discrete Kermack–{McKendrick} Model Adapted to Covid-19},
	volume = {84},
	issn = {1522-9602},
	url = {https://doi.org/10.1007/s11538-022-00994-9},
	doi = {10.1007/s11538-022-00994-9},
	shorttitle = {Back to the Roots},
	pages = {44},
	number = {4},
	journaltitle = {Bulletin of Mathematical Biology},
	shortjournal = {Bull Math Biol},
	author = {Kreck, Matthias and Scholz, Erhard},
	urldate = {2022-09-16},
	date = {2022-02-17},
	langid = {english},
}

@article{ferrari_network_2006,
	title = {Network frailty and the geometry of herd immunity},
	volume = {273},
	url = {https://royalsocietypublishing.org/doi/full/10.1098/rspb.2006.3636},
	doi = {10.1098/rspb.2006.3636},
	pages = {2743--2748},
	number = {1602},
	journaltitle = {Proceedings of the Royal Society B: Biological Sciences},
	author = {Ferrari, Matthew J and Bansal, Shweta and Meyers, Lauren A and Bjørnstad, Ottar N},
	urldate = {2022-09-16},
	date = {2006-11-07},
	note = {Publisher: Royal Society},
}

@article{fine_herd_2011,
	title = {“Herd Immunity”: A Rough Guide},
	volume = {52},
	issn = {1058-4838},
	url = {https://doi.org/10.1093/cid/cir007},
	doi = {10.1093/cid/cir007},
	shorttitle = {“Herd Immunity”},
	pages = {911--916},
	number = {7},
	journaltitle = {Clinical Infectious Diseases},
	shortjournal = {Clinical Infectious Diseases},
	author = {Fine, Paul and Eames, Ken and Heymann, David L.},
	urldate = {2022-09-16},
	date = {2011-04-01},
}

@article{brauer_discrete_2010,
	title = {Discrete epidemic models},
	volume = {7},
	rights = {http://creativecommons.org/licenses/by/3.0/},
	url = {https://www.aimsciences.org/article/doi/10.3934/mbe.2010.7.1},
	doi = {10.3934/mbe.2010.7.1},
	pages = {1},
	number = {1},
	journaltitle = {Mathematical Biosciences \& Engineering},
	author = {Brauer, Fred and Feng, Zhilan and Castillo-Chávez, Carlos},
	urldate = {2022-09-16},
	date = {2010},
	langid = {english},
	note = {Company: Mathematical Biosciences \& Engineering
Distributor: Mathematical Biosciences \& Engineering
Institution: Mathematical Biosciences \& Engineering
Label: Mathematical Biosciences \& Engineering
Publisher: American Institute of Mathematical Sciences},
}

@article{james_use_2021,
	title = {The Use and Misuse of Mathematical Modeling for Infectious Disease Policymaking: Lessons for the {COVID}-19 Pandemic},
	volume = {41},
	issn = {0272-989X},
	url = {https://doi.org/10.1177/0272989X21990391},
	doi = {10.1177/0272989X21990391},
	shorttitle = {The Use and Misuse of Mathematical Modeling for Infectious Disease Policymaking},
	pages = {379--385},
	number = {4},
	journaltitle = {Medical Decision Making},
	shortjournal = {Med Decis Making},
	author = {James, Lyndon P. and Salomon, Joshua A. and Buckee, Caroline O. and Menzies, Nicolas A.},
	urldate = {2022-10-20},
	date = {2021-05-01},
	langid = {english},
	note = {Publisher: {SAGE} Publications Inc {STM}},
}

@article{pastor-satorras_epidemic_2001,
	title = {Epidemic Spreading in Scale-Free Networks},
	volume = {86},
	url = {https://link.aps.org/doi/10.1103/PhysRevLett.86.3200},
	doi = {10.1103/PhysRevLett.86.3200},
	pages = {3200--3203},
	number = {14},
	journaltitle = {Physical Review Letters},
	shortjournal = {Phys. Rev. Lett.},
	author = {Pastor-Satorras, Romualdo and Vespignani, Alessandro},
	urldate = {2022-10-20},
	date = {2001-04-02},
	note = {Publisher: American Physical Society},
}

@article{song_limits_2010,
	title = {Limits of Predictability in Human Mobility},
	volume = {327},
	url = {https://www.science.org/doi/10.1126/science.1177170},
	doi = {10.1126/science.1177170},
	pages = {1018--1021},
	number = {5968},
	journaltitle = {Science},
	author = {Song, Chaoming and Qu, Zehui and Blumm, Nicholas and Barabási, Albert-László},
	urldate = {2022-11-01},
	date = {2010-02-19},
	note = {Publisher: American Association for the Advancement of Science},
}

@article{fontanet_covid-19_2020,
	title = {{COVID}-19 herd immunity: where are we?},
	volume = {20},
	rights = {2020 Springer Nature Limited},
	issn = {1474-1741},
	url = {https://www.nature.com/articles/s41577-020-00451-5},
	doi = {10.1038/s41577-020-00451-5},
	shorttitle = {{COVID}-19 herd immunity},
	pages = {583--584},
	number = {10},
	journaltitle = {Nature Reviews Immunology},
	shortjournal = {Nat Rev Immunol},
	author = {Fontanet, Arnaud and Cauchemez, Simon},
	urldate = {2022-11-01},
	date = {2020-10},
	langid = {english},
	note = {Number: 10
Publisher: Nature Publishing Group},
}

@article{morens_concept_2022,
	title = {The Concept of Classical Herd Immunity May Not Apply to {COVID}-19},
	volume = {226},
	issn = {1537-6613},
	url = {https://europepmc.org/articles/PMC9129114},
	doi = {10.1093/infdis/jiac109},
	pages = {195--198},
	number = {2},
	journaltitle = {The Journal of infectious diseases},
	shortjournal = {J Infect Dis},
	author = {Morens, David M and Folkers, Gregory K and Fauci, Anthony S},
	urldate = {2022-11-01},
	date = {2022-08-01},
	pmid = {35356987},
	pmcid = {PMC9129114},
}

@article{randolph_herd_2020,
	title = {Herd Immunity: Understanding {COVID}-19},
	volume = {52},
	issn = {1074-7613},
	url = {https://www.sciencedirect.com/science/article/pii/S1074761320301709},
	doi = {10.1016/j.immuni.2020.04.012},
	shorttitle = {Herd Immunity},
	pages = {737--741},
	number = {5},
	journaltitle = {Immunity},
	shortjournal = {Immunity},
	author = {Randolph, Haley E. and Barreiro, Luis B.},
	urldate = {2022-11-01},
	date = {2020-05-19},
	langid = {english},
}

@article{lloyd-smith_superspreading_2005,
	title = {Superspreading and the effect of individual variation on disease emergence},
	volume = {438},
	rights = {2005 Nature Publishing Group},
	issn = {1476-4687},
	url = {https://www.nature.com/articles/nature04153},
	doi = {10.1038/nature04153},
	pages = {355--359},
	number = {7066},
	journaltitle = {Nature},
	author = {Lloyd-Smith, J. O. and Schreiber, S. J. and Kopp, P. E. and Getz, W. M.},
	urldate = {2022-11-10},
	date = {2005-11},
	langid = {english},
	note = {Number: 7066
Publisher: Nature Publishing Group},
}

@misc{ozbay_parameterizing_2022,
	title = {Parameterizing Network Graph Heterogeneity using a Modified Weibull Distribution},
	url = {http://arxiv.org/abs/2212.06994},
	doi = {10.48550/arXiv.2212.06994},
	number = {{arXiv}:2212.06994},
	publisher = {{arXiv}},
	author = {Ozbay, Sinan A. and Nguyen, Maximilian M.},
	urldate = {2022-12-20},
	date = {2022-12-13},
	eprinttype = {arxiv},
	eprint = {2212.06994 [physics, stat]},
}

\renewcommand{\thefigure}{S\arabic{figure}}
\setcounter{figure}{0}

\section*{Methods}
\subsection*{Construction of Contact Networks With A Specified Amount of Heterogeneity}

We generate the graph as follows. First, we select a value of $\sigma \in (0,1)$, with $0$ representing a highly homogeneous degree distribution and $1$ representing a highly heterogeneous degree distribution. The value of $\sigma$ chosen maps to a unique value of $\alpha \in (0,\infty)$ where $\alpha$ is the shape parameter of the Weibull distribution, through the transformation $$\sigma = e^{-\alpha} $$

Next, we fix $\lambda$, which can be thought of as the number at which the degree distribution will be centered at. Simulations in the Results section are conducted on graphs of 1000 nodes and 5000 connections. Simulations on graphs several orders of magnitude larger produce results that are similar. 

The scale parameter $\lambda$ and shape parameter $\sigma$ specify a unique 2-parameter Weibull distribution, given by:
\begin{equation}
f(x; \lambda, \sigma) = \frac{-ln(\sigma)}{\lambda}(\frac{x}{\lambda})^{-ln(\sigma) - 1} e^{-(x/\lambda)^{-ln(\sigma)}};   x\ge 0; \sigma \in (0,1) ; \lambda > \mathbb{R}^+  \label{eqn:pdf}
\end{equation}

We then sample from this distribution to achieve the desired network size. With the sample obtained, we construct the graph using the configuration model. This specifies a graph $G(\sigma)$ which is sampled from the space of all graphs that are compatible with the degree distribution generated per the above procedure.

\subsection*{Simulation of a Discrete-Time SIR Model on A Graph}
The model of the simulation is as follows. Given a graph $G(\sigma)$ representing the contact network of interest, fix a transmission probability $\tau$ and recovery probability $\gamma$:
\begin{enumerate}
\item At time $t_0$, fix a small fraction $f$ of nodes to be chosen uniformly on the graph and assign them to the Infected state. The remaining $(1 - f)$ fraction of nodes start as Susceptible. 
\item For each $i \in [1,T]$, for each pair of adjacent S and I nodes, the susceptible node becomes infected with probability $\tau$.
\item For each $i \in [1,T]$, each infected node recovers with probability $\gamma$.
\item At time $T$, record two quantities: The final attack rate and the herd immunity threshold at the peak of the epidemic.
\item Repeat steps (1-4) n = 150 times for each value of $\tau$.
\item Repeat steps (1-5) for each value of $\sigma$.
\end{enumerate}

\subsection*{Discretizing the SIR Ordinary Differential Equations Model}
To numerically compute the trajectory of the deterministic system of ODEs, we discretize the Kermack-McKendrick SIR model and then apply Euler's method. The SIR equations are given as:
\begin{align}
\frac{dS}{dt} &= - \beta S(t) I(t)\\
\frac{dI}{dt} &= \beta S(t) I(t) - \gamma I(t)\\
\frac{dR}{dt} &= \gamma I(t)
\end{align}

Commonly, the Euler method is used when numerically integrating these equations. For an ordinary differential equation of the form $\frac{du}{dt} = f(u)$ this amounts to computing an approximate solution $\tilde u(t)$ according to
\begin{align}
\tilde u(t_n) = \tilde u(t_{n-1}) + f(\tilde u(t_{n-1})) \ \Delta t
\end{align}

In the case of the SIR model we can write:
\begin{align}
\frac{\tilde S(t_n) - \tilde S(t_{n-1})}{\Delta t} = -\beta \tilde S(t_{n-1}) \tilde I(t_{n-1}) 
\end{align}
and similarly for $\tilde I$ and $\tilde R$, where tilde $(\sim)$ denotes an approximate numerical solution.

In Figure S1, we show that the oscillations are a direct result of time-discretization by discretizing the SIR ODE model with an increasingly small time-step and measure the HIT attained across these deterministic simulations of epidemics. 

As we can see, finer time steps lead to smaller oscillations in the HIT as we vary the transmissibility parameter $\beta$. Thus, as the discretization of time becomes negligible for the ODE model, the oscillations disappear.

\section*{Acknowledgements}
The authors would like to acknowledge the members of the Levin Lab for their suggestions and feedback. S.A.O. was supported by the Global Health Program at Princeton University. B.F.N. was supported by the Carlsberg Foundation under its Semper Ardens programme (grant CF20-0046).

\section*{Author Contributions}
S.A.O. and M.M.N. designed and performed research. All authors analyzed the results and wrote the paper. 

\section*{Additional Information}
The authors declare no competing interests.

\section*{Data Availability}
The datasets and code generated during and/or analysed during the current study are available from the corresponding author on reasonable request.

\end{document}


\begin{center}
\textbf{\large Supplemental Materials: Bifurcations in the Herd Immunity Threshold for Discrete-Time Models of Epidemic Spread}
\end{center}

\renewcommand{\thefigure}{S\arabic{figure}}
\subsection*{Figure \ref{fig:timestep}. HIT as a function of time step size in difference equations.}
\begin{figure}[h]
    \centering
    \includegraphics[width=0.32\textwidth]{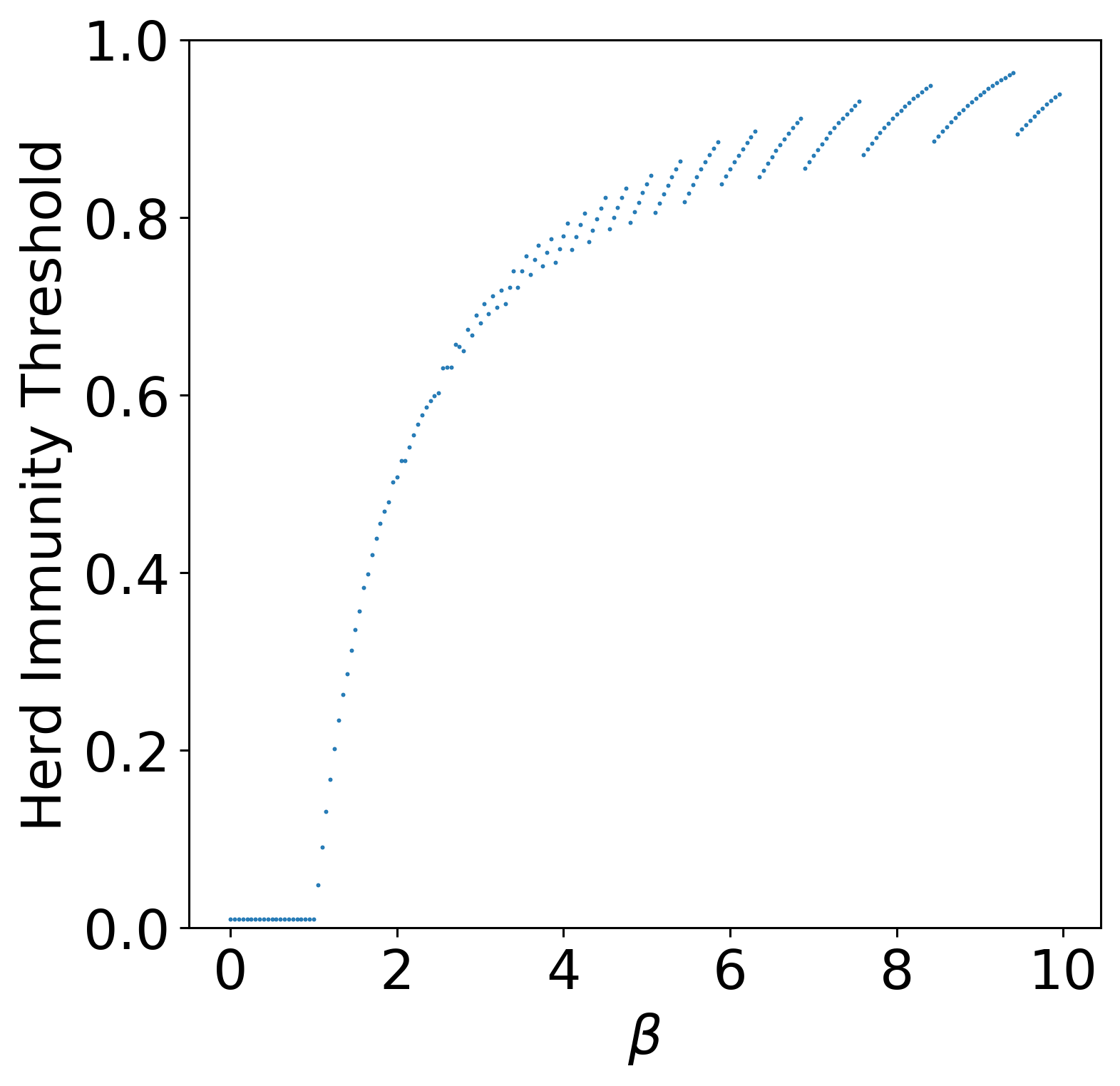}
    \includegraphics[width=0.32\textwidth]{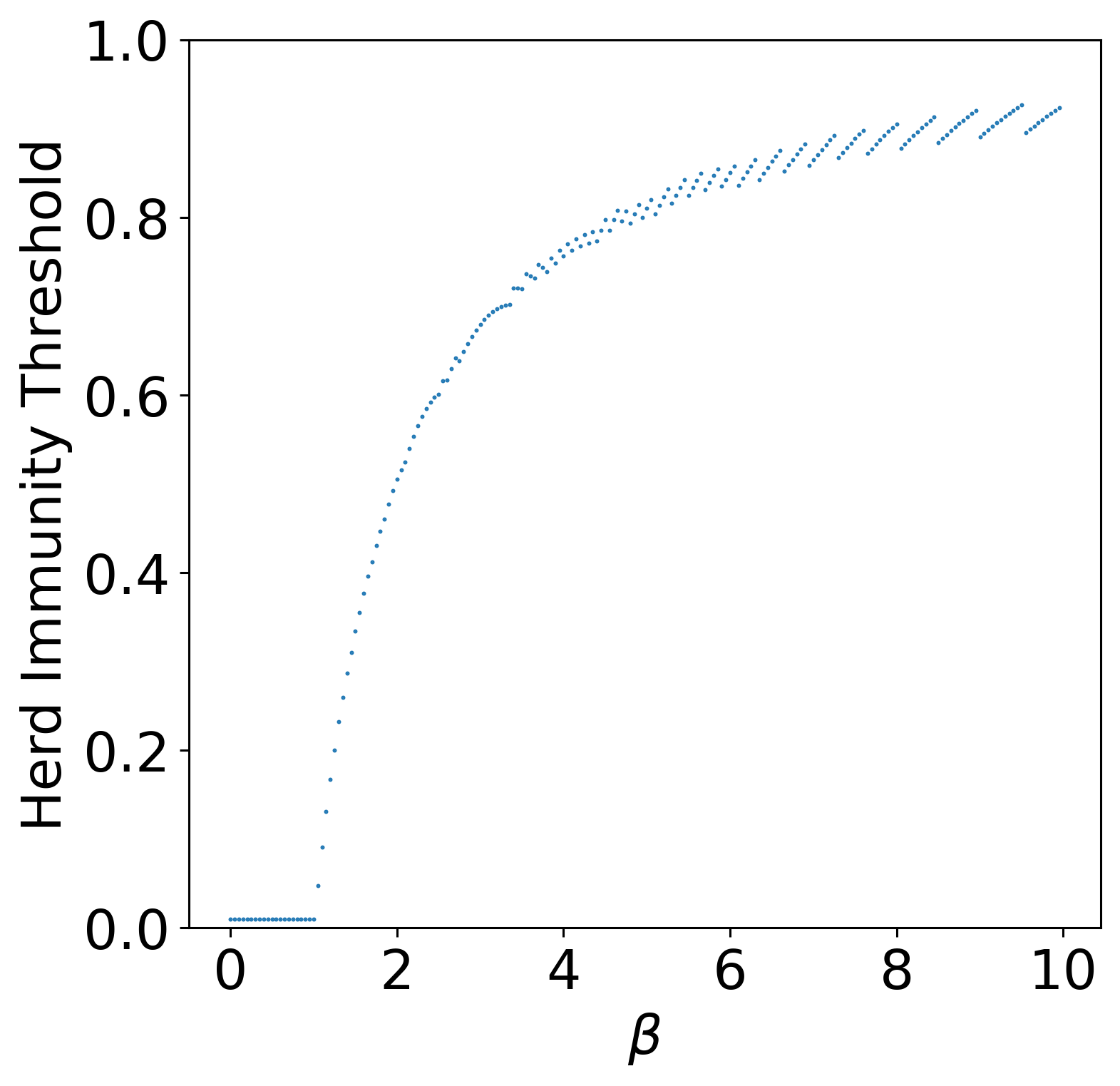}
    \includegraphics[width=0.32\textwidth]{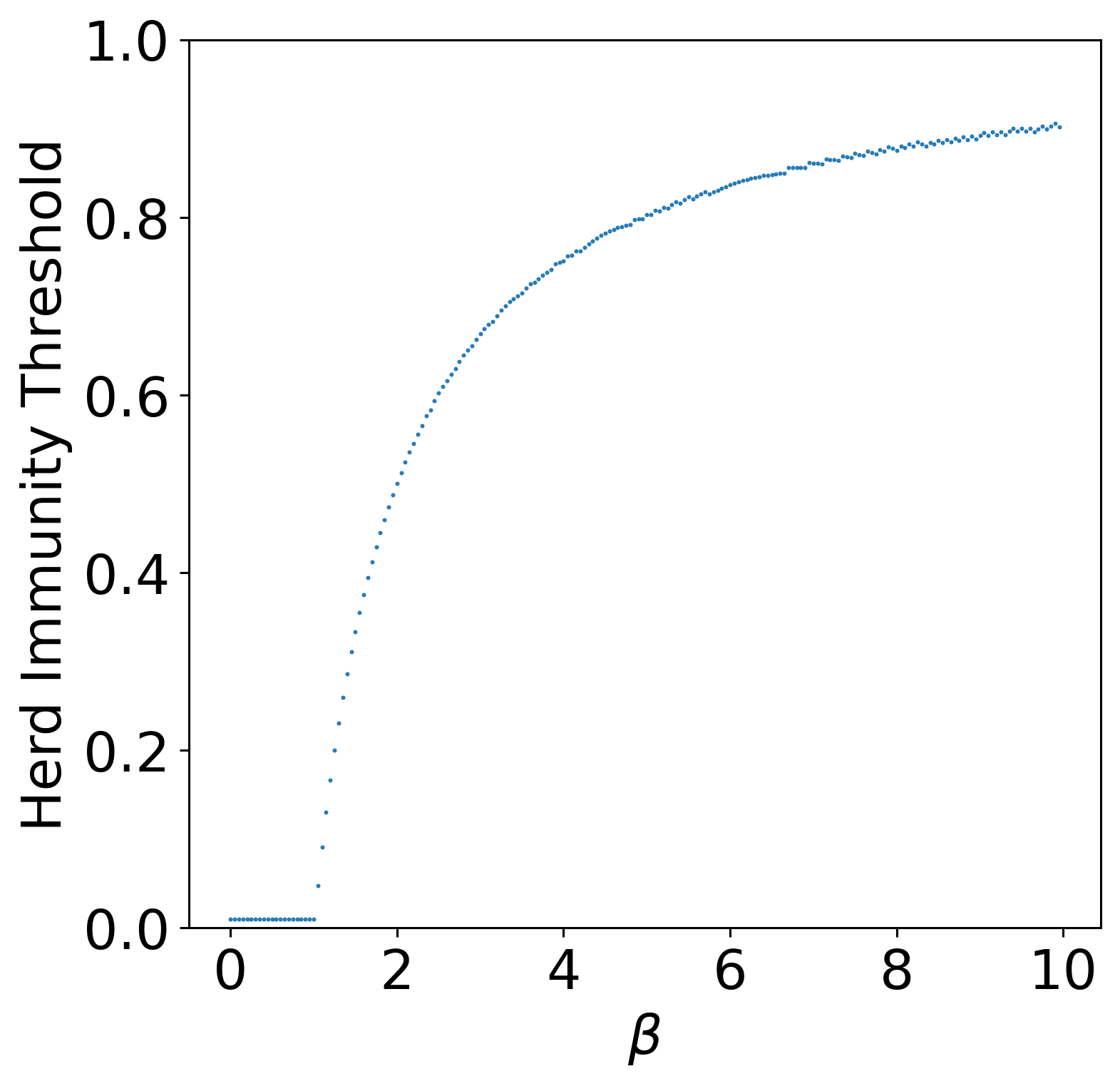}
    \caption{HIT for various values of the transmissibility parameter $\beta$, measured for deterministic ODE models of SIR for $dt = [0.1, 0.05, 0.01]$} \label{fig:timestep}
\end{figure}